# Thermodynamic properties of CrMnFeCoNi high entropy alloy at elevated electronic temperatures

Nikita Medvedev[1,2,*]

1) Institute of Physics, Czech Academy of Sciences, Na Slovance 1999/2, 182 00 Prague 8, Czech Republic
2) Institute of Plasma Physics, Czech Academy of Sciences, Za Slovankou 3, 182 00 Prague 8, Czech Republic

## Abstract

The Cantor alloy (equiatomic CrMnFeCoNi) is a high-entropy alloy with unique physical properties and radiation resistance. To model its response to intense laser pulses, the parameters of the electronic ensemble are required. In this work, the electronic heat capacity, thermal conductivity, and electron-phonon coupling strength at elevated electronic temperatures are evaluated using a combined approach that incorporates tight-binding molecular dynamics and the Boltzmann equation. The damage threshold fluence is estimated for a wide range of photon energies, from XUV to hard X-rays. It is found that at the electronic temperatures ~24,000 K (absorbed dose ~6 eV/atom), the Cantor alloy experiences *nonthermal* melting due to modification of the interatomic potential induced by electronic excitation, even without the increase of the atomic temperature. This effect must be included in reliable models of CrMnFeCoNi ablation under ultrafast laser irradiation.

## Keywords:

Cantor alloy; CrMnFeCoNi high-entropy alloy; nonthermal melting; electron-phonon coupling; electron heat conductivity; electron heat capacity.

## I. Introduction

The equiatomic CrMnFeCoNi alloy, also known as the Cantor alloy, is an archetypical example of high-entropy alloys (HEA) [1,2]. Its unique properties attracted attention in a wide range of practical applications. Its high ductility, fracture toughness, strain hardening, high resistance to wear, corrosion, and hydrogen embrittlement make it suitable for its applications in harsh environments [3–5]. The recyclability and low fabrication costs of the high entropy alloys make them a promising candidate for wide industrial uses [6,7].

In particular, the Cantor alloy and its derivatives are known for their radiation hardness [5,8,9]. It attracted much attention as a candidate material for nuclear applications and radiation-loaded conditions [9]. However, most of the previous studies included low-dose-rate (low-flux) irradiation scenarios, except for the laser ablation of high entropy alloys [10]. High-dose-rate irradiation may produce qualitatively different effects and material states [11,12].

[*] Corresponding author: email: nikita.medvedev@fzu.cz, ORCID: 0000-0003-0491-1090

The laser micro-machining of materials is a commonly used industrial technique to design and manufacture materials with desired properties [13,14]. It involves precise material melting and ablation after high-intensity irradiation, forming sub-micron features on the surface and in the depth of the material. Pulsed-laser irradiation of HEA requires an understanding of the fundamental processes taking place in highly excited states under ultrafast energy deposition [15,16].

Generally speaking, ultrafast laser-matter interaction takes place via a sequence of processes, starting with photoabsorption by the electronic ensemble of the irradiation material. It drives the electronic system out of equilibrium, inducing the electron cascades of secondary ionisations, thermalisation among the electrons, and energy exchange with the atomic system (typically known as the electron-phonon coupling) [16]. Atomic heating via this coupling may lead to atoms acquiring sufficiently high temperatures, eventually overcoming their potential energy barriers and undergoing a phase transition – e.g., thermal melting [15]. Additionally, at high electronic temperatures (i.e., high intensities of the impinging laser pulse), some materials may exhibit *nonthermal* melting: a modification of the interatomic potential leading to lattice instabilities even without atomic heating [17,18]. The nonthermal melting is well established in semiconductors and insulators, but its occurrence in metals remains a subject of ongoing research, and is primarily supported by recent computational studies [19].

The thermodynamic and transport properties in the excited electronic system of matter are crucial parameters for understanding ultrafast laser ablation [16]. These parameters are hard to find for the Cantor alloy due to its complex atomic structure – simulation boxes with a large number of atoms are required to sample the random placement of various atoms. To fill this knowledge gap, we perform a detailed study of the equiatomic Cantor alloy with the help of the XTANT-3 combined simulation tool [20]. It integrates tight-binding molecular dynamics, transport Monte Carlo, and Boltzmann collision integral methods into one model with feedback, delivering a state-of-the-art simulation method. It allows for the study of coupled effects of thermal, nonthermal, and nonequilibrium kinetics in the irradiated HEA.

## II. Model

XTANT-3 code applied to simulation material parameters evolution under laser irradiation [20]. The model combines the following approaches to trace the effects of irradiation: the transport Monte Carlo method and the Boltzmann equation to trace the electronic system, and tight-binding molecular dynamics for the atomic system.

The photoabsorption, the induced nonequilibrium electron kinetics, and the Auger decays of produced core holes are modelled with the event-by-event individual particle Monte Carlo approach [21]. The photoabsorption cross sections, the ionisation potentials of core shells, and Auger-decay times are used from the EPICS2023 database [22] (see Appendix). The electron impact ionisation cross section is calculated in the framework of the linear response theory (the complex-dielectric function formalism) with the single-pole method [23] (the calculated electron inelastic mean free paths are also shown in the Appendix).

Slow electrons populating the bottom of the conduction band are modelled with the Boltzmann equation, including the electron-electron and electron-phonon interactions [24]. The evolution of the electronic energy levels (band structure) is traced with the transferrable tight-binding method [25–27]. The $sp^3d^5$-based PTBP density-functional tight binding parametrisation is employed [28]. It includes pairwise interaction of all elements, allowing for modelling of complex materials, such as the Cantor alloy. The diagonalisation of the electronic Hamiltonian produces the electronic energy levels (molecular orbitals) and the transient interatomic forces [25]. Since the changes in the electronic distribution function directly affect the interatomic potential, the method is capable of describing the nonthermal melting [27,29].

The electron heat capacity is evaluated as the derivative of the electronic entropy at the given electron temperature; the electron chemical potential is calculated numerically on the transient electronic energy levels obtained from the TB module [30]. The electron heat conductivity is obtained with the help of the Onsager coefficients, including the electron-phonon and electron-electron contributions (on the *k*-vector grid of 7x7x7 points in the supercell) [31]. The electron-phonon coupling parameter is calculated *via* the nonperturbative dynamical-coupling approach [32].

For a reliable evaluation of the electronic parameters, a sufficiently large supercell is required – here, 320 atoms in the simulation box are modelled [32]. The atoms are placed randomly on the fcc grid and allowed to thermalise at room temperature for a few hundred femtoseconds prior to the simulation of irradiation. To eliminate artifacts of a particular placement of atoms, the evaluated electron heat capacity, conductivity, and the electron-phonon coupling parameters are averaged over 200 independent realizations with different random placement of atoms in the supercell, initial velocities (according to the Maxwellian distribution at room temperature), and parameters of the electronic temperature increase (see details in Ref. [32]).

The molecular dynamic simulations use a time-step of 0.1 fs for evaluation of the electron-phonon coupling parameter, and a 1 fs step for radiation-damage simulations. Martyna-Tuckerman's 4th-order algorithm is employed for propagation of atomic trajectories [33].

The simulation box size of 14.76 x 14.48 x 18.45 $Å^3$ is defined *via* the steepest descent algorithm, producing the material density of 7.557 $g/cm^3$, which is slightly lower than the previously reported values of ~7.9 $g/cm^3$ [34], which is typical for multi-element transferrable DFTB simulations.

All the illustrations of the atomic snapshots are prepared with the help of OVITO [35].

## III. Results

### *1. Electronic properties*

We start with the evaluation of the electronic density of states (DOS) in the Cantor alloy, see Figure 1 showing a reasonable comparison with the previously reported DOS calculated with the density functional theory in Ref. [36]. The total and partial DOS are shown, resolving the contribution of *s*, *p*, and *d* orbitals from each element. The bottom of the conduction band below the Fermi level is mostly formed by the d-orbitals of Ni, Co, Fe, and Mn atoms (in the order of the energy depth), whereas the DOS above the Fermi level is mainly formed by the d-orbitals of Cr.

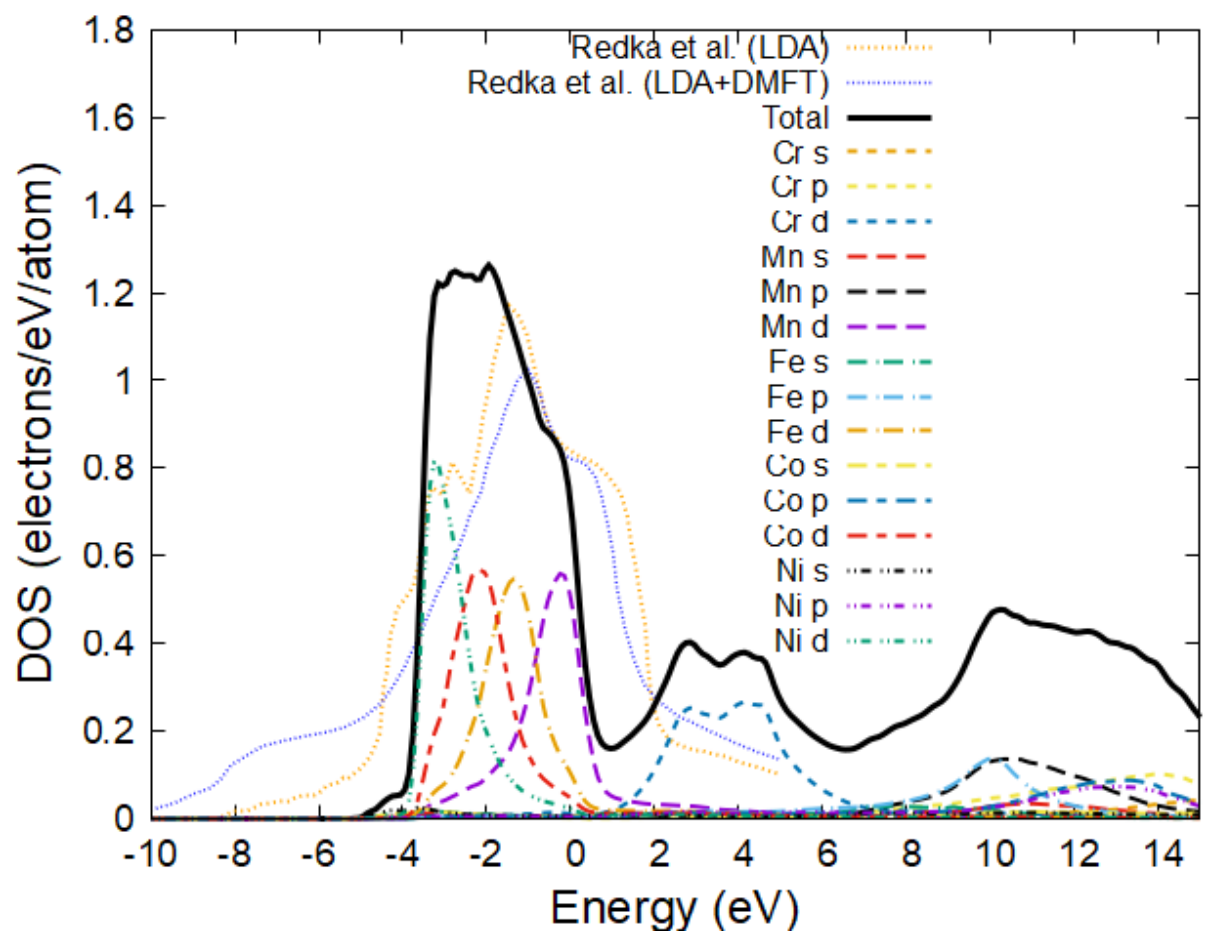

*Figure 1. Electronic density of states in fcc CrMnFeCoNi calculated with XTANT-3. Total and partial DOS are shown, counted from the Fermi energy. For comparison, DFT calculations within LDA and LDA+DMFT by Redka et al. are shown* [36].

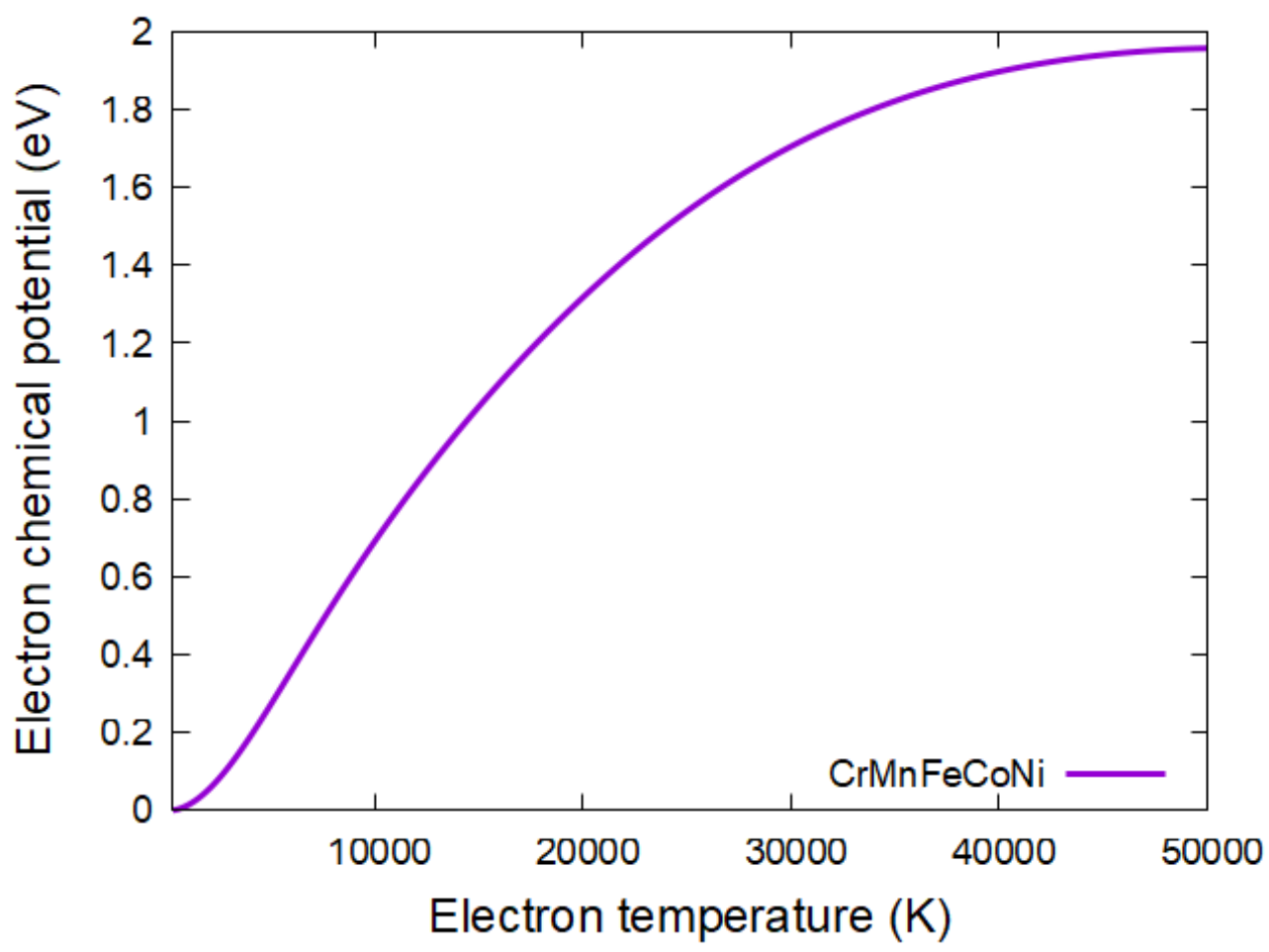

*Figure 2. Electron chemical potential in fcc CrMnFeCoNi calculated with XTANT-3.*

The electronic chemical potential rises monotonously from the Fermi level with an increase in the electronic temperature, Figure 2. Having the DOS and the electronic chemical potential, the electronic heat capacity can be calculated [37]. It is shown in Figure 3 that the electronic heat capacity rises significantly up to the electronic temperatures ~30,000 K and starts to decrease after that.

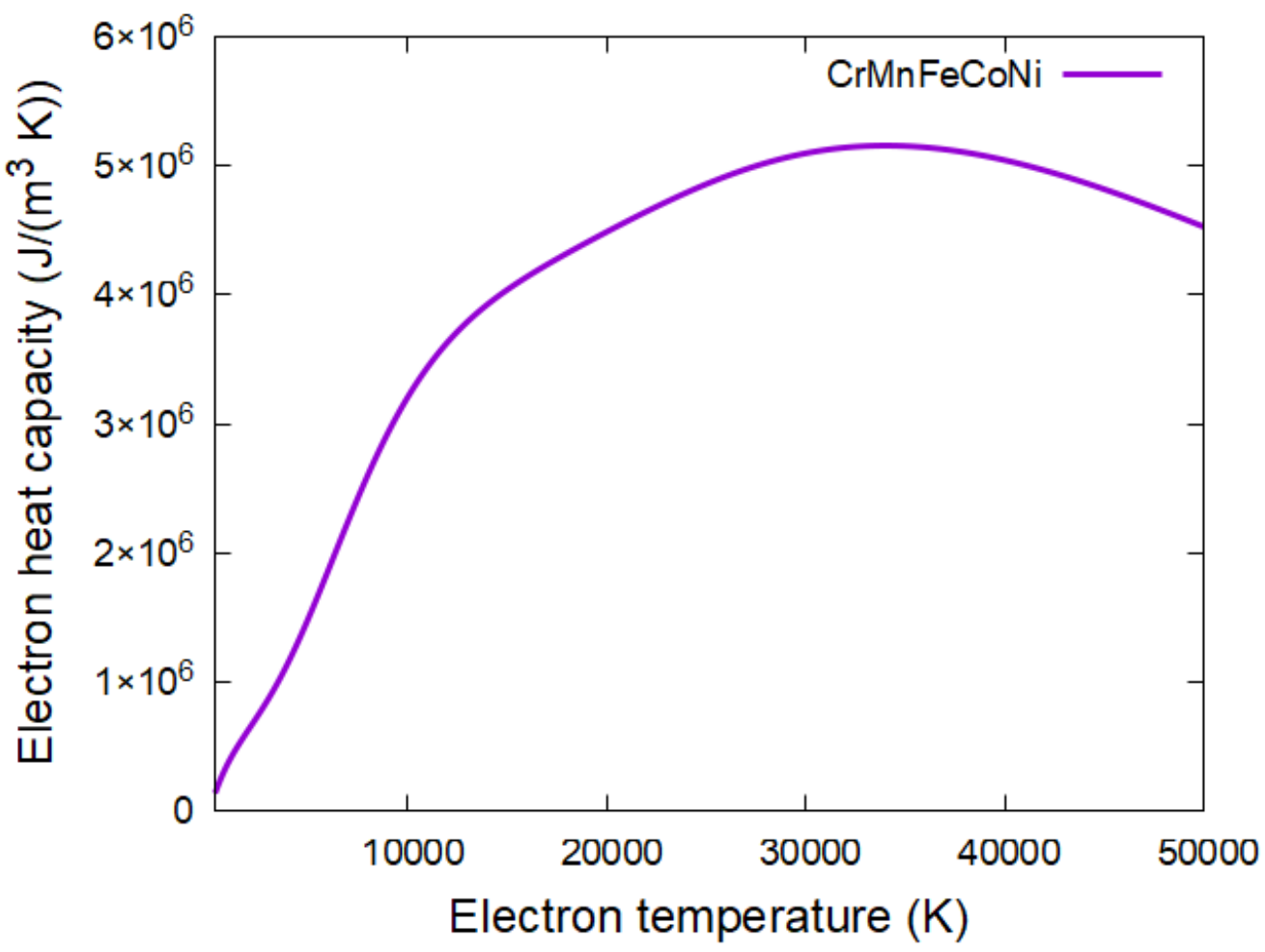

*Figure 3. Electron heat capacity in fcc CrMnFeCoNi calculated with XTANT-3.*

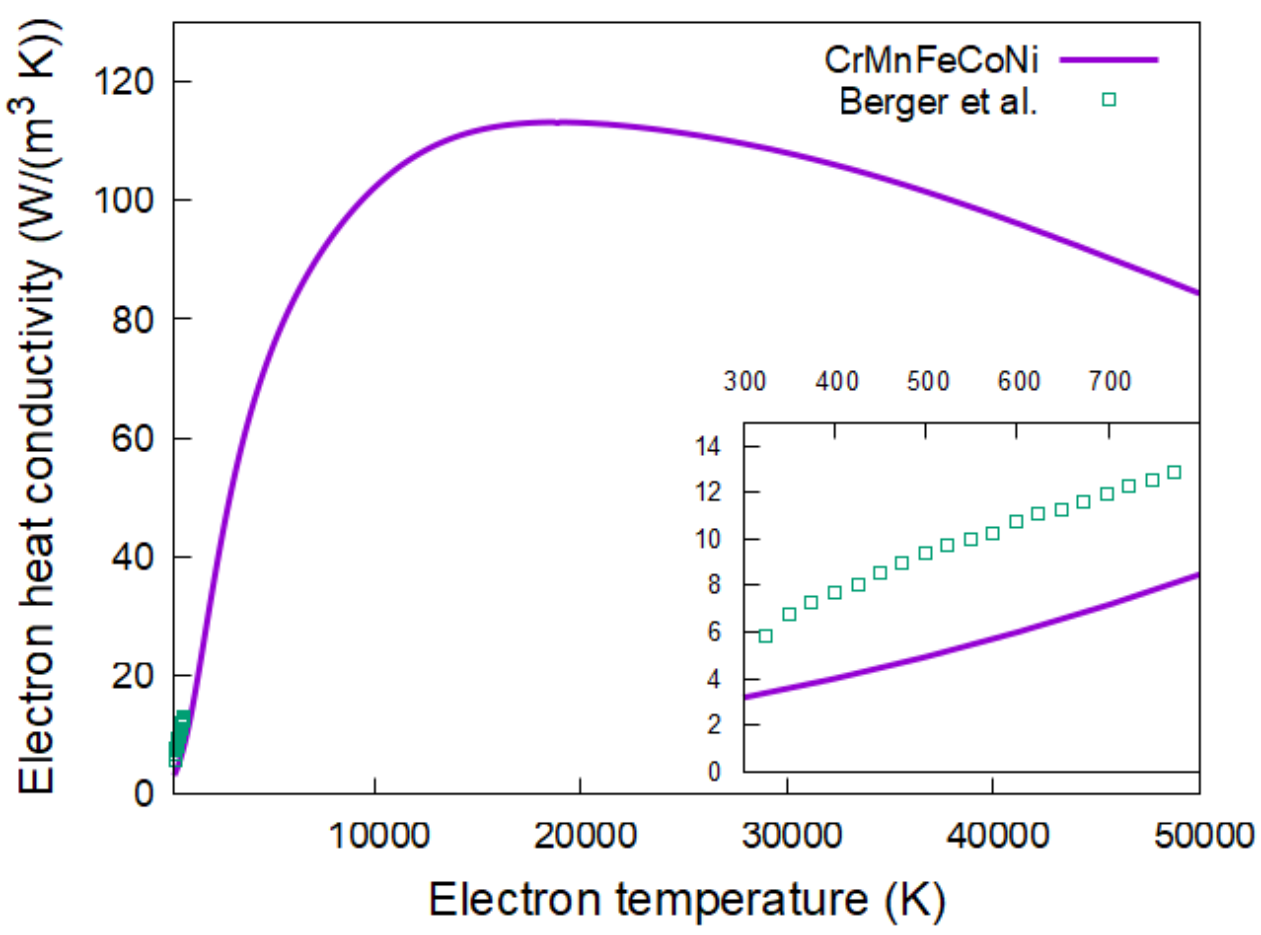

*Figure 4. Electron heat conductivity in fcc CrMnFeCoNi calculated with XTANT-3, compared with low-temperature measurements by Berger et al.*[38] *. The inset zooms into the low-temperature region for better visibility of the comparison.*

The electronic heat conductivity calculated with XTANT-3 is presented in Figure 4, compared with the experimental low-temperature limit [38]. Berger *et al*. used the measured resistivity to calculate the electron thermal conductivity via the Wiedemann-Franz law [38]. An overall reasonable agreement is observed in Figure 4, with the XTANT-3 calculations following the increasing trend with the increase in the electron temperature up to ~18,000 K, followed by a slow decrease. The electronic heat conductivity in the Cantor alloy is dominated by the electron-phonon contribution, whereas the electron-electron contribution is negligible. The electronic heat conductivity is significantly smaller than that of the constituent elemental metals [31], and also than, e.g., in stainless steel [39].

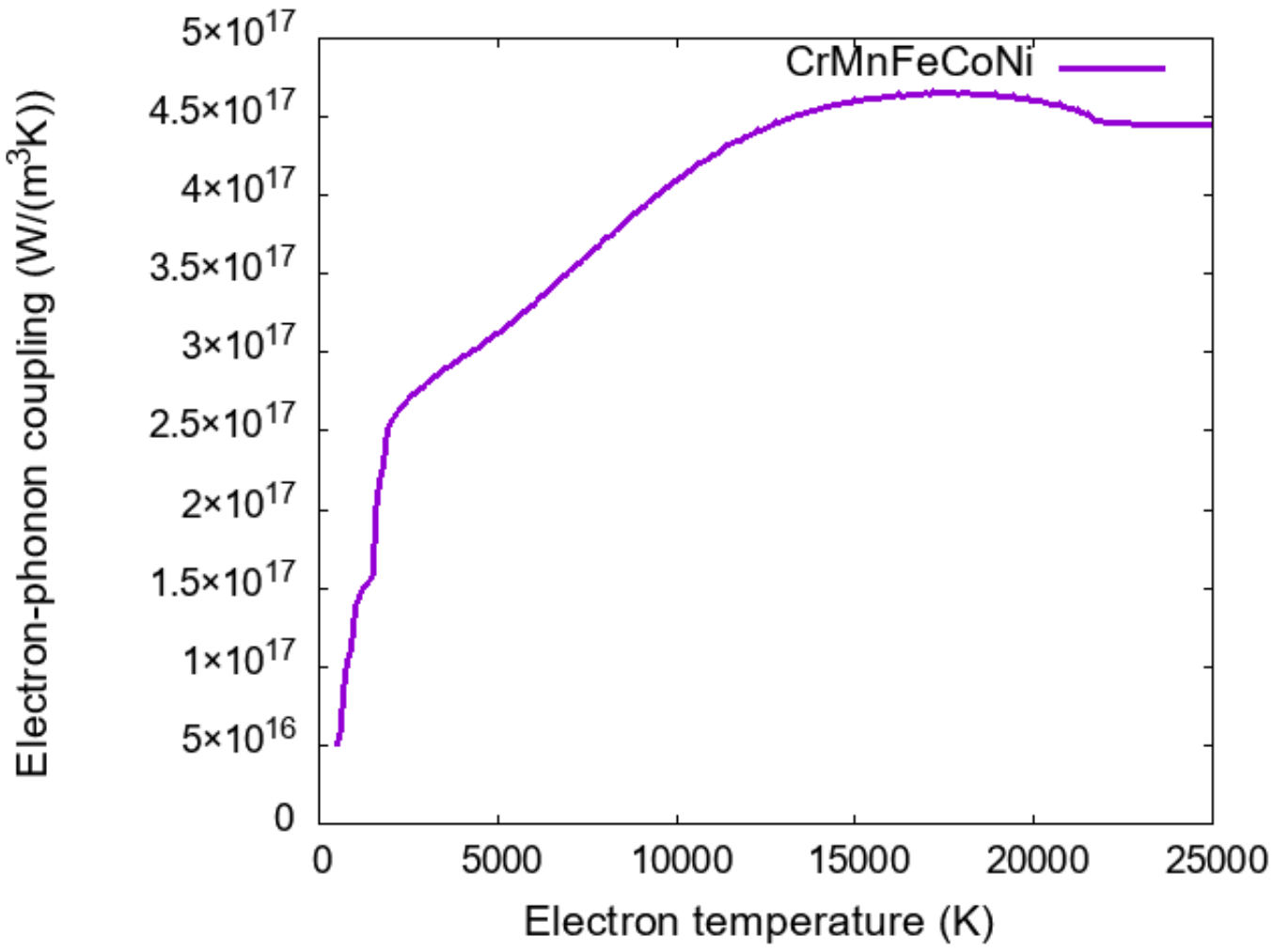

*Figure 5. Electron-phonon (electron-ion) coupling in fcc CrMnFeCoNi calculated with XTANT-3.*

The calculated electron-phonon coupling parameter is shown in Figure 5. The coupling strength in CrMnFeCoNi is relatively high, comparable to that in cobalt, chromium, and iron, especially at high electron temperatures [32].

## *2. Dynamical effects*

A series of dynamical simulations of irradiation of the Cantor alloy with various doses was performed, identifying the onset of material damage. Melting in CrMnFeCoNi takes place above the deposited dose of ~0.3-0.4 eV/atom. An example of the atomic snapshots after irradiation with 1 eV/atom dose, 30 eV photon energy and 10 fs (full width at half maximum, FWHM) pulse duration is shown in Figure 6.

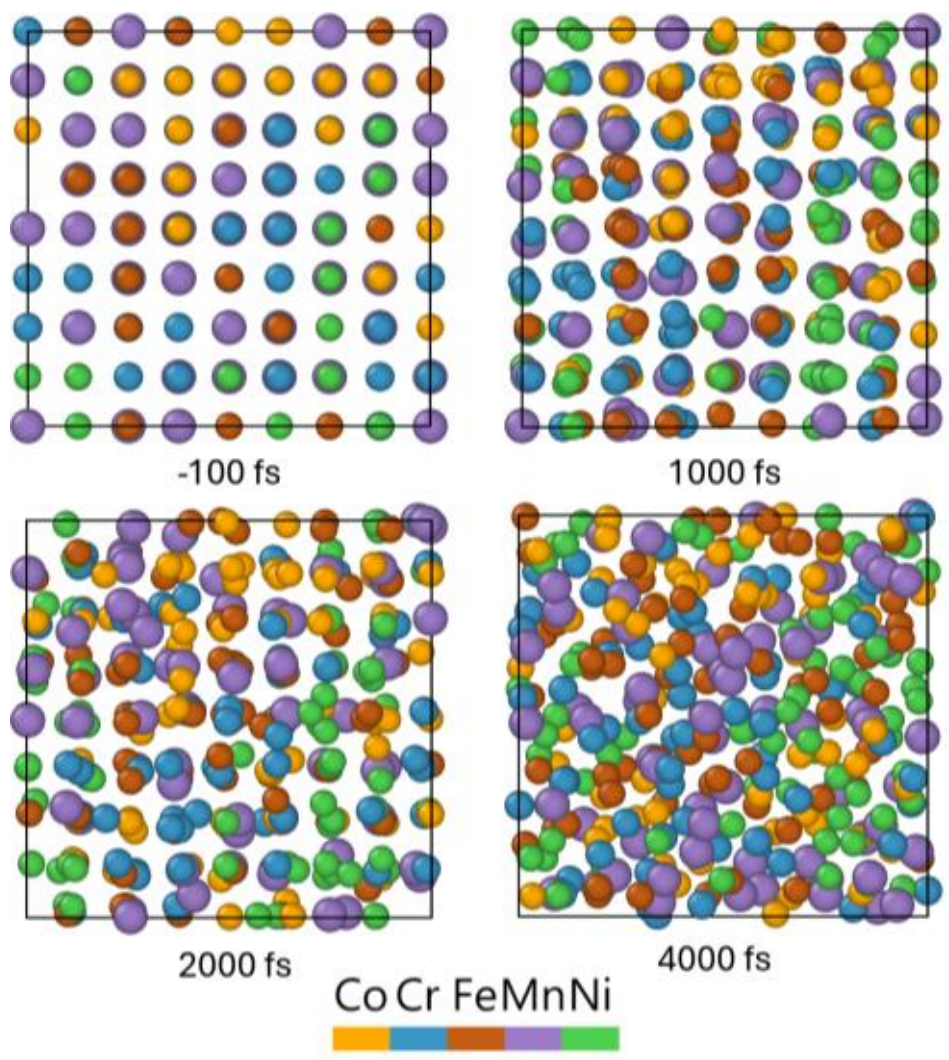

*Figure 6. Atomic snapshots of CrMnFeCoNi irradiated with a laser pulse of 30 eV photons, 10 fs FWHM duration, 1 eV/atom absorbed dose; simulation including electron-phonon coupling (nonadiabatic dynamics).*

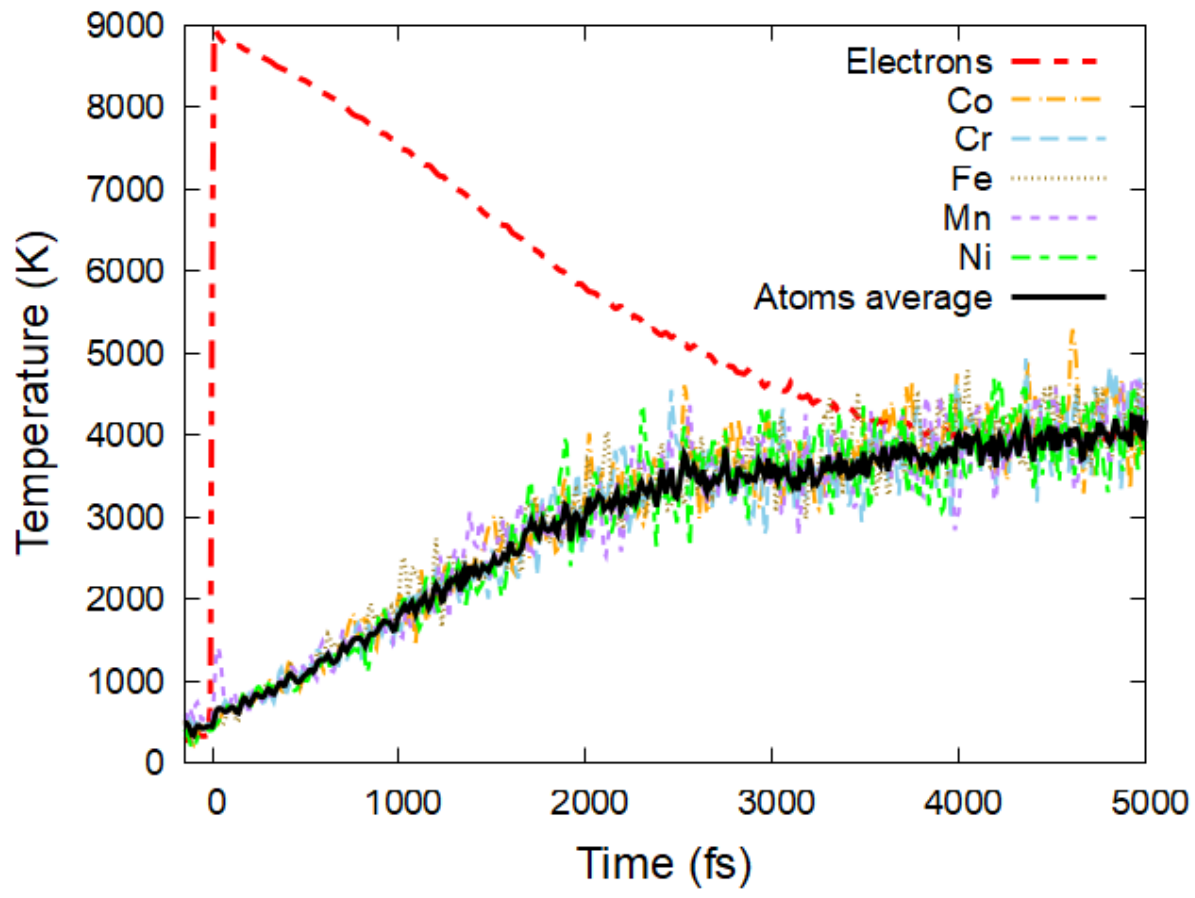

*Figure 7. Electronic and atomic temperatures (total and element-specific) in CrMnFeCoNi irradiated with a laser pulse of 30 eV photons, 10 fs FWHM duration, 1 eV/atom absorbed dose; simulation including electron-phonon coupling (nonadiabatic dynamics).*

The experimental melting temperature is reported to be around ~1563 K – 1613 K [40]. At the deposited dose of 1 eV/atom, this temperature is overcome already at sub-picosecond timescales, see Figure 7. The thermal melting starts to take place due to atomic heating by the excited electronic system *via* the electron-phonon coupling (Figure 5). By the time of 3-4 ps, the atomic system disorders completely (cf. Figure 6).

Interestingly, the manganese sublattice experiences transient acceleration immediately after the rise in the electronic temperature due to photoabsorption (see a spike in the Mn temperature around 0 fs in Figure 7). This selective nonthermal acceleration of atoms suggests that the interatomic potential is changed noticeably by the electronic excitation [41]. The same effect of the Mn subsystem reaction to electronic excitation was observed in stainless steel simulations [39].

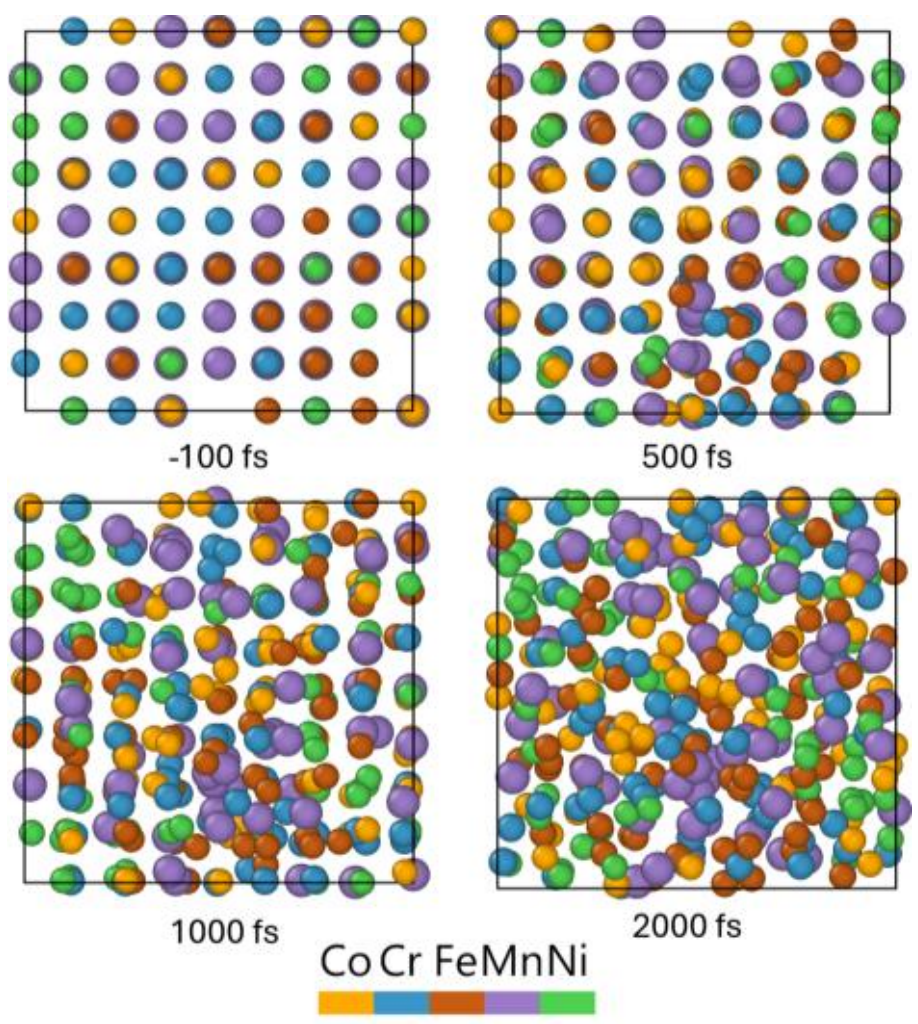

*Figure 8. Atomic snapshots of Born-Oppenheimer simulation (no electron-phonon coupling) of CrMnFeCoNi irradiated with a laser pulse of 30 eV photons, 10 fs FWHM duration, 7 eV/atom absorbed dose.*

To evaluate the possibility of nonthermal melting, a separate series of simulations with various deposited doses in the Born-Oppenheimer (BO) approximation is performed (excluding the electron-phonon coupling). It is found that the atomic lattice loses stability at the deposited doses above ~6 eV/atom (the electronic temperature ~24,000 K) in the BO simulation, see an example in Figure 8. This figure demonstrates that the atomic lattice in the Cantor alloy can disorder purely due to modification of the interatomic potential induced by the electronic excitation, without atomic heating. In contrast to most covalent materials [17,27,42], the nonthermal melting is not ultrafast but occurs at ~ 1 ps timescales, comparable to the thermal melting timescales.

It is worth emphasising that the BO simulation is only performed to show the existence of such an effect as purely nonthermal melting in high entropy alloys; in practice, ultrafast energy deposition often leads to intertwined thermal and nonthermal effects [43,44]. An experimental validation is needed, although it is a nontrivial task to discern the two effects [45].

It is also interesting to note that the nonthermal melting threshold in the Cantor alloy is much higher than that in, e.g., stainless steel, another complex metallic alloy [39]. This finding is in line with the notion of the exceptional radiation resistance of the CrMnFeCoNi high entropy alloy [5,8,9].

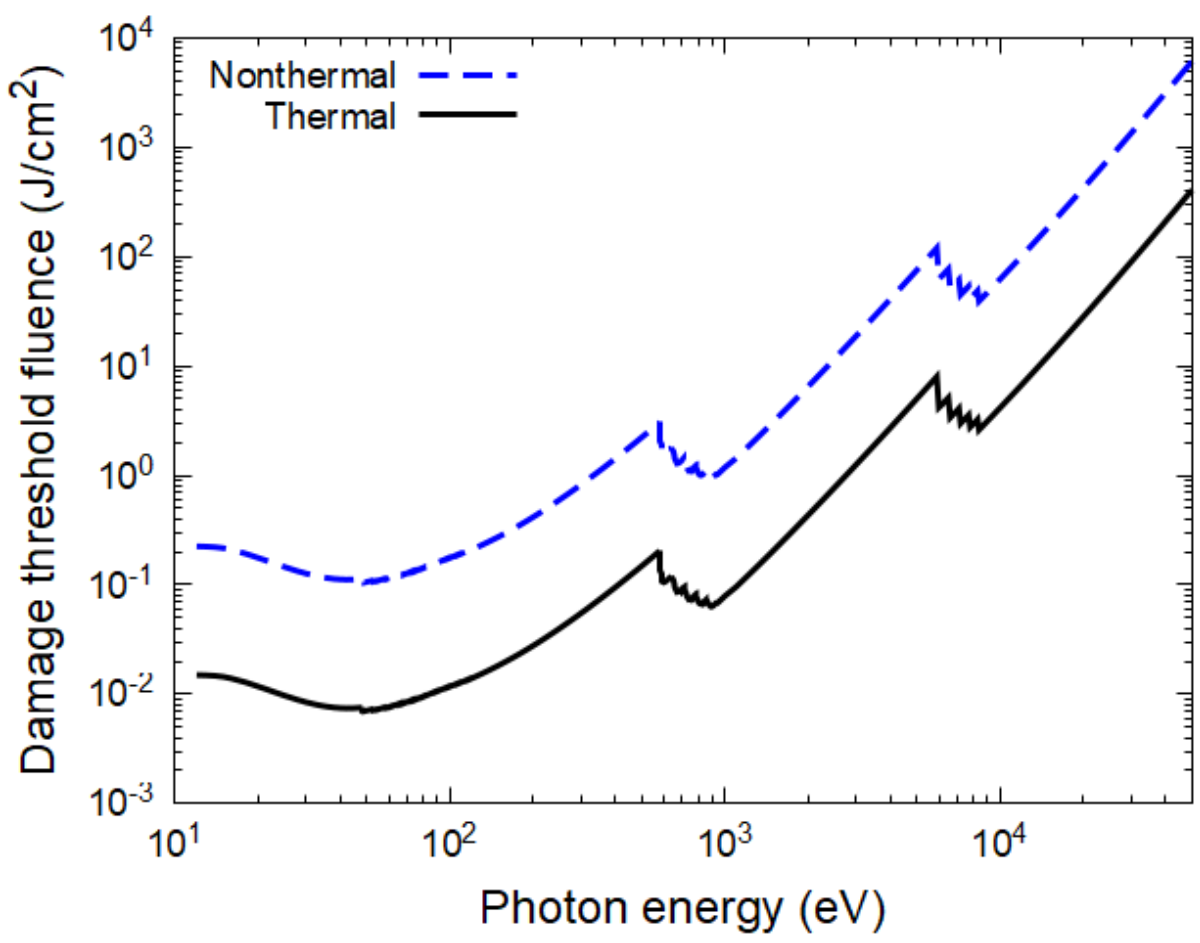

*Figure 9. Thermal and nonthermal melting threshold fluences in CrMnFeCoNi estimated from the damage doses predicted with XTANT-3.*

Assuming normal photon incidence and no particle and energy transport, the damage threshold dose may be converted into the incident fluence [27]. The threshold fluences for thermal and nonthermal melting in the Cantor alloy in a broad range of photon energies are shown in Figure 9. This estimate may be used to guide the experiments and application of laser-irradiation of CrMnFeCoNi.

## IV. Conclusions

Equiatomic CrMnFeCoNi high entropy alloy (the Cantor alloy) is simulated with the help of the XTANT-3 hybrid code. The electronic heat capacity and conductivity are calculated up to the electronic temperatures of ~50,000 K. The electron-phonon coupling parameter is evaluated using the nonperturbative dynamical coupling approach up to the electronic temperature of ~20,000 K. The thermal damage threshold is estimated as ~0.3-0.4 eV/atom, where material melting occurs at the picosecond timescale. The damage threshold fluence as a function of the incident photon energy up to hard X-rays is estimated.

Our simulations indicate that the Cantor alloy may exhibit *nonthermal* melting, inducing atomic disorder due to changes in the interatomic potential triggered by the electronic excitation. The atomic lattice loses stability at the deposited doses above ~6 eV/atom (the electronic temperature ~24,000 K) in a Born-Oppenheimer simulation (excluding the electron-phonon coupling); including electron-phonon coupling (nonadiabatic simulation) leads to intertwining of the two effects. The manganese sublattice is especially susceptible to nonthermal acceleration due to electronic excitation, exhibiting a transient temperature rise. These effects must be considered in modelling and interpreting the experiments on laser ablation of the high entropy alloys.

## V. Conflicts of interest

There are no conflicts to declare.

## VI. Data and code availability

The code XTANT-3 used to obtain electronic properties and simulate irradiation effects, and input data including photon and electron mean free paths, are available from [20]. The tables with the calculated electronic heat capacity, conductivity, and electron-phonon coupling parameter are available from [46].

## VII. Acknowledgements

Computational resources were provided by the e-INFRA CZ project (ID:90254), supported by the Ministry of Education, Youth and Sports of the Czech Republic.

## VIII. Appendix

The total and partial photon attenuation lengths in the Cantor alloy, constructed from the EPICS2023 database accounting for the stoichiometry of the material, are shown in Figure 10 (left panel) [22]. The electronic inelastic mean free paths in the same material, calculated with the linear response theory using the single-pole approximation, accounting for plasmon scattering [23], are shown in Figure 10 (right panel).

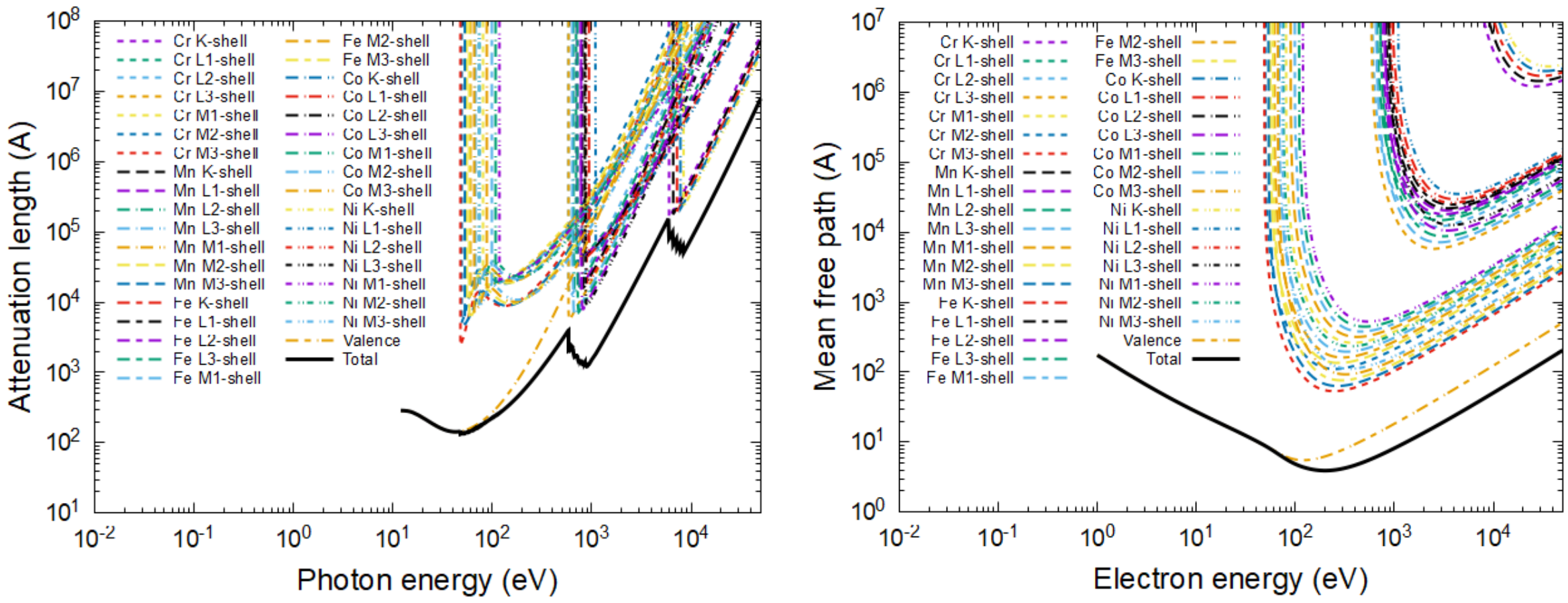

*Figure 10. Photon attenuation length (left panel) and electronic inelastic mean free path (right panel) in CrMnFeCoNi. Total values are shown in black solid lines. Partial ones, for each atomic shell of each element, are dashed and coloured.*

# IX. Funding statement declaration

The author thanks the financial support from the Czech Ministry of Education, Youth, and Sports (grant nr. LM2023068).